%% file: paper.tex
\documentclass[a4paper,11pt]{amsart}

	\usepackage{amsmath,amssymb,amsthm,amsaddr}
	\usepackage{mathtools}
	\usepackage{xspace}
	\usepackage{xcolor,graphicx}
	\usepackage{hyperref}
	\hypersetup{colorlinks=true,linktocpage=true,citecolor=green!80!red,linkcolor=green!20!red,urlcolor=blue!50!red}
	\usepackage[small]{caption}
	\allowdisplaybreaks

	\usepackage[a4paper]{geometry}

	\usepackage{booktabs}

	\input{newcommands}

	\title[]{Universality of the topological string at large~radius and NS-brane resurgence}
	\author{Ricardo Couso-Santamar\'ia}
	\email{santamaria@math.tecnico.ulisboa.pt}
	\address{CAMGSD, Departamento de Matem\'atica, \\ Instituto Superior T\'ecnico, Universidade de Lisboa,\\ Av. Rovisco Pais 1, 1049-001 Lisboa, Portugal}

\begin{document} 

	\begin{abstract}
	We show that there is a natural universal limit of the topological string free energies at the large radius point. 
	The new free energies keep a nonholomorphic dependence on the complex structure moduli space and their functional form is the same for all Calabi--Yau geometries, compact and noncompact alike. 
	The asymptotic nature of the free energy expansion changes in this limit due to a milder factorial growth of its coefficients, and this implies a transseries extension with instanton effects in $\exp(- 1/g_s^2)$, of NS-brane type, rather than $\exp(-1/g_s)$, of D-brane type. 
	We show a relation between the instanton action of NS-brane type and the volume of the Calabi--Yau manifold which points to a possible interpretation in terms of NS5-branes. 
	A similar rescaling limit has been considered recently leading to an Airy equation for the partition function which is here used to explain the resurgent properties of the rescaled transseries. 
	\end{abstract}
	\maketitle
	\tableofcontents

\newpage

\section{Introduction}

	Topological string theory is a perturbatively defined theory of maps from Riemann surfaces to a Calabi--Yau manifold. 
	The perturbative\footnote{We include a superindex ${}^{(0)}$ to specify that a quantity is related to perturbation theory.} free energy $F^{(0)}=\sum_g g_s^{2g-2}\,F^{(0)}_g$ is an asymptotic, divergent, series in the string coupling constant $g_s$, due to ${F^{(0)}_g\sim (2g)!}$ as $g\to\infty$.   
	Such factorial growth is an example of a commonplace phenomenon which in string theory was first described in \cite{Gross:1988ib,shenker1990}. 
	The coefficients of $F^{(0)}$, namely $F^{(0)}_g(z,\overline{z})$ are nonholomorphic functions of the complex structure moduli of the Calabi--Yau, that we denote collectively as $z$. 
	Here we are looking at the free energy from the B-model side; in the A-model they depend on the K\"ahler structure of the mirror geometry. 
	An efficient approach to computing $F^{(0)}$ is by means of the holomorphic anomaly equations \cite{Bershadsky:1993ta,Bershadsky:1993cx} using a finite set of nonholomorphic generators called propagators ($S^{zz}$, $S^{z}$, $S$, $K_z$) instead of the complex conjugate of the modulus, $\overline{z}$. 
	The resulting free energies are then polynomials \cite{Yamaguchi:2004bt,Alim:2007qj} in these generators. 
	See \cite{Alim:2012gq} for a nice review of these topics. 

	The asymptotic nature, and intrinsically perturbative definition, of the topological string free energy calls for a nonperturbative completion. 
	There has been recent work on this direction that culminated in a conjecture for such a completion based on the spectral theory of an operator that quantizes the geometry of toric Calabi--Yau manifolds, \cite{Grassi:2014zfa} (see also \cite{Marino:2015nla} for a recent review). 
	A different approach, which partly inspires the present work, is based on the theory of resurgence \cite{Ecalle:1981uy} (see \cite{sauzin:2014aa} for a mathematical review and \cite{Marino:2012zq} for a more physical perspective). 
	The idea in this case is to provide a formal series that goes beyond perturbation theory and that includes nonanalytic terms including factors such as $\rme^{-A/g_s}$, which can be regarded physically as instanton corrections of some kind, with $A$ the instanton action. 
	These formal series go by the name of \emph{transseries}. 
	The final step for the nonperturbative completion involves the resummation of the transseries which necessarily requires additional physical input to be uniquely defined. 
	If the transseries is \emph{resurgent} the coefficients that build it satisfy very constraining relations. 
	In fact, it is expected that all the information of the transseries can be extracted from perturbation theory alone, if computational power is not a restriction.  
	This was the approach set up in \cite{cesv13,Couso-Santamaria:2014iia}. 
	The transseries found there had the schematic form
	\begin{equation}
		F = \sum_{g=0}^\infty g_s^{2g-2} \, F^{(0)}_g(z,\overline{z}) + \sum_{n=1}^\infty \s_\text{D}^n\,\rme^{-n A_\text{D}(z)/g_s}\, \sum_{g=0}^\infty g_s^g \, F^{(n)}_g(z,\overline{z}),
		\label{eq:Fusualtransseries}
	\end{equation}
	where the various instanton actions, $A_\text{D}(z)$, are holomorphic and the other coefficients are not. 
	The transseries parameters, $\sigma_\text{D}$, that make \eqref{eq:Fusualtransseries} into a family of solutions, and the instanton actions are labeled by D to denote that the instanton effects are of D-brane type. 
	The instanton actions $A_\text{D}(z)$ have a fundamental role in the theory of resurgence: they control how the coefficients $F^{(0)}_g$ grow factorially with $g$, namely
	\begin{equation}
		F^{(0)}_g \sim \frac{(2g)!}{A_\text{D}^{2g}}, \quad \text{ as $g\to\infty$},
	\end{equation}
	and they materialize as singularities of the Borel transform of $F^{(0)}(g_s)$,
	\begin{equation}
		\CB[F^{(0)}](s) = \sum_{g=0}^\infty \frac{F^{(0)}_g}{(2g)!}s^{2g}.
		\label{eq:CBF0s}
	\end{equation}
	If the set of singularities is discrete and we can extend the Borel transform analytically along any path that avoids them then $F^{(0)}(g_s)$ is called resurgent. 
	Since the singularities are responsible for the divergence of the series, the absence of singularities should indicate that the series is actually convergent. 

	It was noticed in \cite{Couso-Santamaria:2014iia}, for the Calabi--Yau geometry of local $\BP^2$, that when the complex modulus $z$ goes to $0$, that is, to the large radius point, the instanton actions move off to infinity. 
	However the resulting free energy series in that limit keeps showing asymptotic behavior, although of slightly different nature. 
	The growth of the coefficients $F^{(0)}_g\sim (2g)!/A_\text{D}^{2g}$ disappears when the instanton actions are infinite and subleading contributions are revealed growing like $g!$. 
	We also find that the effect of the large radius limit is universal, that is, the same for all Calabi--Yau geometries, both local and compact, so it is not restricted to local $\BP^2$.  
	We show in this work that this universal $g!$-growth is linked to a nonperturbative effect of the form
	\begin{equation}
		\rme^{- A_\text{NS}/g_s^2},
	\end{equation}
	which is familiar from NS-branes in string theory. 
	In fact we show that $A_\text{NS}$ is related, in the large radius limit, to the volume of the Calabi--Yau geometry, which suggests a connection with NS5-branes. 
	The main question that must be answered is what is the physical interpretation of this new nonperturbative effect intrinsic to all topological string theories? 

	In Section~\ref{sec:factorialdivergenceatthelargeradiuspoint} we explain what exactly happens at the large radius point, how a rescaling is needed to make sense of it, and how the result is independent of the Calabi--Yau geometry, compact or local. 
	We make contact with the recent work \cite{Alim:2015qma} where a similar rescaling was introduced and show their close relation. 
	Section~\ref{sec:propertiesoftherescaledfreeenergies} shows the universality of the rescaled free energies using the holomorphic anomaly equations. 
	We describe the large-order growth, transseries extension, and resurgent properties of the free energies helped by the connection with the Airy equation found in \cite{Alim:2015qma}. 
	Section~\ref{sec:closedstringinstantoncontributionstothetopologicalstring} widens the scope from the large radius point back to the full moduli space and considers transseries sectors involving $\rme^{-A_\text{NS}/g_s^2}$. 
	We describe $A_\text{NS}$ and its relation to the Calabi--Yau volume, and speculate about its possible physical interpretation in terms of NS-branes.
	We end with the conclusions.

\section{Factorial divergence at the large radius point}
\label{sec:factorialdivergenceatthelargeradiuspoint}

	\subsubsection*{The Borel plane for local $\BP^2$}

		One of the results of \cite{Couso-Santamaria:2014iia} was the identification of instanton actions that contribute to the nonperturbative behavior of the (B-model) topological string free energy on the mirror of local $\BP^2$. 
		These instanton actions are, as anticipated in \cite{Drukker:2011zy}, periods of the geometry, and in particular holomorphic functions of the complex structure modulus $z$. 
		From the point of view of resurgence the instanton actions are incarnated as poles of the Borel transform of $F^{(0)}$, defined by \eqref{eq:CBF0s}.  
		The actions in the Borel plane change with the complex structure modulus but not with the propagator.
		It was noticed in \cite{Couso-Santamaria:2014iia} that when we approach the large radius point, $z=0$, all the actions move off to infinity, leaving the Borel plane empty of singularities.\footnote{The constant map contribution to the free energies is present for every geometry and it implies a constant instanton action, $4\pi^2\rmi$. We subtract the constant map contribution to get rid of this action without losing generality. }

		The  naive expectation would be that at the large radius point the free energy would become convergent because no poles would drive the usual $(2g)!$-growth. 
		But this is not what was found. 
		First of all, a naive limit $z\to 0$ with a fixed value of the propagator $S^{zz}$ would make the free energies blow up due to the Yukawa coupling (see equation \eqref{eq:localP2Czzz}). 
		If on the other hand we let the propagator acquire its holomorphic value in the large radius frame
		\begin{equation}
			S^{zz}_{\hol,\text{LR}} = z^2\left( \frac{1}{2} + 9z - 54z^2 + \ldots \right)
			\label{eq:localP2Szzhol}
		\end{equation}
		then the free energies vanish when $z\to0$. 
		(They would become the pure constant map contribution but we have removed it.) 
		The trivial vanishing series is indeed convergent but not very interesting. 
		The natural way forward is to rescale the propagator,
		\begin{equation}
			S^{zz} =: z^2 \, \Sigma,
			\label{eq:Szzz2Sigmarescaling}
		\end{equation}
		and define a new set of free energies
		\begin{equation}
			H^{(0),\BP^2}_g(\Sigma) := \lim_{z\to 0} F^{(0)}_g (z, S^{zz} = z^2 \Sigma), \qquad g\geq2. 
			\label{eq:Hbf0gdef}
		\end{equation}
		The free energies for $g=0,1$ do not follow the same pattern. 
		The relevant genus $0$ information is captured by the Yukawa coupling which is holomorphic and undergoes no rescaling. 
		The limit of $F^{(0)}_1$ is considered later in equation \eqref{eq:H01HAE}. 

		It was found in \cite{Couso-Santamaria:2014iia} that the new rescaled perturbative free energy
		\begin{equation}
			H^{(0),\BP^2} = \sum_{g=2}^\infty g_s^{2g-2}\,H^{(0),\BP^2}_g,
			\label{eq:Hu0P2}
		\end{equation}
		once all instanton actions associated to the $(2g)!$-growth have disappeared from the Borel plane, is still asymptotic. 
		The coefficients now grow factorially like
		\begin{equation}
			H^{(0),\BP^2}_g \sim \frac{\G(g-1)}{A_H(\Sigma)^{g-1}}, \quad \text{as $g\to\infty$},
		\end{equation}
		for a particular function $A_H(\Sigma)$. 
		Two important points to mention: the new instanton action that controls the growth is not holomorphic and the growth itself goes like $g!$ rather than $(2g)!$. 
		This growth is invisible from the Borel transform \eqref{eq:CBF0s} and it can only be discovered at $z=0$. 
		We will comment on these issues and their significance later in the paper. 
		Before that let us explore the rescaling \eqref{eq:Szzz2Sigmarescaling} a little further.

	\subsubsection*{The rescaling at the large radius point}

		The free energies $F^{(0)}_g(z,S^{zz})$ are polynomial in the propagator and rational in the modulus. 
		When we rescale $S^{zz}$ by $z^2$ and take the large radius limit we capture some of the coefficients of these rational functions. 
		For example, for local $\BP^2$ and genus $g=2$, we have
		\begin{equation}
			F^{(0)}_2 = C_{zzz}^2 \left( \frac{5}{24} (S^{zz})^3 - \frac{3 z^2}{16} (S^{zz})^2 + \frac{z^4}{16} S^{zz} -  \frac{(11 -162 z -729 z^2)z^6}{1920} \right) - \frac{1}{1920}, 
		\end{equation}
		where the last term removes the constant map contribution and $C_{zzz}$ is the Yukawa coupling,
		\begin{equation}
			C_{zzz} = -\frac{1}{3z^3 (1+27z)}.
			\label{eq:localP2Czzz}
		\end{equation}
		The power $z^{-6}$ in $C_{zzz}^2$ is canceled by other similar powers of $z$ that accompany each term $\Sigma^k$, $k=0,1,2,3$, and we find
		\begin{equation}
			H^{(0),\BP^2}_2 = \left(-\frac{1}{3}\right)^2 \left( \frac{5}{24}\Sigma^3 - \frac{3}{16} \Sigma^2 + \frac{1}{16}\Sigma - \frac{11}{1920} \right) - \frac{1}{1920}.
		\end{equation}
		It is worth noting that the coefficient of the ambiguity captured in the rescaling, namely $-11/1920$, is the one fixed by the constant-map boundary condition at $z=0$, not by the gap condition at the conifold locus. 

		Since the rescaled free energies vanish in the holomorphic limit, it is natural to introduce a variable $X:= \Sigma - \s_\hol$, where $\s_\hol=\frac{1}{2}$ is the holomorphic limit of $\Sigma$ in the example of local $\BP^2$, see \eqref{eq:localP2Szzhol}. 
		Then,
		\begin{equation}
			H^{(0),\BP^2}_2 = \left(-\frac{1}{3}\right)^2 X \left( \frac{5}{24}X^2 + \frac{1}{8}X + \frac{1}{32} \right).
		\end{equation}
		For generic genus $g$ we find $H^{(0),\BP^2}_g = X^{2g-3}\Pol(X;g)$. 
		The total degree of $H^{(0),\BP^2}_g$ in $X$ equal to $3g-3$ is a generic property of $F^{(0)}_g$ \cite{Alim:2007qj} that is preserved in the large radius limit. 

		The existence of a large radius point is a generic feature of all Calabi--Yau manifolds that corresponds to a large value of the K\"ahler structure moduli in the A-model mirror geometry. 
		It is then natural to apply the same limit to other geometries, starting with toric ones having a single modulus. 
		The resulting rescaled free energies differ from those of local $\BP^2$ in the particular coefficients. 
		However it is not difficult to find the right variable $u$, proportional to $X$, that leads to universal results for all $g$. 
		Indeed, if we define $u$ by
		\begin{equation}
			S^{zz} =: z^2 \left( \s_\hol + \tilde{b}\k^{-1}u \right)
			\label{eq:Szzu}
		\end{equation}
		then the rescaled free energies are proportional to universal polynomials $H^{(0),u}_g(u)$,
		\begin{equation}
			H^{(0),\text{geom}}_g = \left( \tilde{b}^3 \k^{-1} \right)^{g-1} \, H^{(0),u}_g.
			\label{eq:HgeomHu}
		\end{equation}
		Here $\k$ is the classic intersection number and $\tilde{b}$ is related to a topological invariant of the Calabi--Yau, see \eqref{eq:tildebb}.  
		Note that the prefactor can be combined with $g_s^{2g-2}$ in \eqref{eq:Hu0P2}. 
		We show this universality property of the rescaling in the next section. 
		We have checked \eqref{eq:HgeomHu} for various toric geometries, such as local del Pezzo, but also for a compact geometry, the mirror quintic, and for an example with two moduli, local $\BP^1\times\BP^1$, see Section~\ref{sec:universalityoftherescaling}. 

		The first few free energies $H^{(0),u}_g(u)$ are
		\begin{align}
			H^{(0),u}_2 &= u \left( \frac{5}{24}u^2 + \frac{1}{2}u + \frac{1}{2} \right), 
			\label{eq:H0u2}\\ 
			H^{(0),u}_3 &= u^3 \left( \frac{5}{16}u^3 + \frac{5}{8}u^2 + \frac{1}{2}u + \frac{1}{6} \right), \\ 
			H^{(0),u}_4 &= u^5 \left( \frac{1105}{1152}u^4 + \frac{15}{8}u^3 + \frac{25}{16}u^2 + \frac{2}{3}u + \frac{1}{8} \right), \\ 
			H^{(0),u}_5 &= u^7 \left( \frac{565}{128}u^5 + \frac{1105}{128}u^4 + \frac{15}{2}u^3 + \frac{175}{48}u^2 + u + \frac{1}{8} \right), \\ 
			H^{(0),u}_6 &= u^9 \left( \frac{82825}{3072}u^6 + \frac{1695}{32}u^5 + \frac{12155}{256}u^4 + 25u^3 + \frac{525}{64}u^2 + \frac{8}{5}u + \frac{7}{48} \right).
			\label{eq:H0u6}
		\end{align}
		The leading coefficients, let us call them $a^{(0)}_g$, have been discussed recently in \cite{Alim:2015qma}. 
		In that work they appear as the coefficient of the highest degree term in the propagator $S^{zz}$, that is
		\begin{equation}
			F^{(0)}_g = a^{(0)}_g \, C_{zzz}^{2g-2} \left( S^{zz} \right)^{3g-3} + \ldots
			\label{eq:F0ga0gldots}
		\end{equation}
		To obtain the generating series $\sum_g \l_s^{2g-2} \, a^{(0)}_g$ the following rescaling was considered
		\begin{equation}
			S^{zz} \mapsto \varepsilon^{2/3}\, S^{zz}, \quad g_s \mapsto \varepsilon^{-1}\, g_s,
			\label{eq:AYZrescaling}
		\end{equation}
		along with $\l_s^2 := g_s^2 C_{zzz}^2 (S^{zz})^3$ and $\varepsilon\to0$.  
		A natural question formulated in \cite{Alim:2015qma} asked for a physical interpretation of the parameter $\varepsilon$. 
		In light of the previous discussion it is natural to consider the relation
		\begin{equation}
			\varepsilon  \leftrightarrow z^{-3}.
		\end{equation}
		This implies the rescaling $S^{zz} \mapsto z^{-2} \, S^{zz}$ (l.h.s. of \eqref{eq:AYZrescaling}) that we considered, with a different notation, in \eqref{eq:Szzz2Sigmarescaling}. 
		Note, however, that $z$ still goes to zero, not infinity. 
		The $\varepsilon \to 0$ limit, designed to capture the coefficients $a^{(0)}_g$, is based on the polynomial structure of the free energies alone (lower degree terms in \eqref{eq:F0ga0gldots} vanish in the limit) and does not involve the complex modulus $z$ at all. 
		The large radius limit, while doing the same rescaling of the propagator, does not rescale $g_s$. 
		As a consequence lower order terms in \eqref{eq:F0ga0gldots} also contribute to $H^{(0),u}_g(u)$. 
		The generating series for the coefficients $a^{(0)}_g$ can be obtained from $H^{(0),u}$ as
		\begin{equation}
			\lim_{u\to\infty} \sum_{g=2}^\infty g_s^{2g-2} \, H^{(0),u}_g(u) = \sum_{g=2}^\infty \l_s^{2g-2} \, a^{(0)}_g
		\end{equation}
		where $g_s^2 u^3 = \l_s^2$ is kept finite. 

		In this way the $\varepsilon$-limit of \cite{Alim:2015qma} can be geometrically interpreted as a large radius limit and, moreover, it can be refined to keep nonholomorphic information. 
		The large radius limit is also universal once the proper variables are considered, as in \eqref{eq:Szzu} and \eqref{eq:HgeomHu}. 
		Note also that the rescaling by $z^2$ is natural from the point of view of the holomorphic limit \eqref{eq:localP2Szzhol}. 
		It was also implicitly considered in \cite{Alim:2013eja} when constructing a set of generators instead of the propagators that would behave nicely under (generalized) modular properties; see also \cite{Alim:2012ss}. 
		In local $\BP^2$, for example, the second Eisenstein series $\hat{E}_2$ can take the role of the propagator $S^{zz}$ \cite{Aganagic:2006ho,Haghighat:2008gw} with the property that its holomorphic limit is $\CO(z^0)$ rather than $\CO(z^2)$ as in \eqref{eq:localP2Szzhol}, so the factor $z^2$ in front of $S^{zz}$ has naturally been taken care of.

\section{Properties of the rescaled free energies}
\label{sec:propertiesoftherescaledfreeenergies}

	In this section we present some of the properties of the rescaled free energies $H^{(0),u}_g$: 
	\begin{itemize}
		\item they are universal, that is, the same for every geometry, local or compact, 
		\item they satisfy a rescaled version of the holomorphic anomaly equations and also an equation in $g_s$ that can be cast into Airy form, as in \cite{Alim:2015qma},
		\item there is a transseries extension to the perturbative series whose resurgent properties are captured by those of the Airy function and whose factorial growth is subleading, to all orders, to the standard $(2g)!$-growth familiar to string theories.
	\end{itemize}

	\subsection{Universality of the large radius limit}
	\label{sec:universalityoftherescaling}

		To simplify the discussion we restrict ourselves to local geometries with one modulus for which we can keep $S^{zz}$ and set the other generators $S^{z}$, $S$ and $K_z$ safely to their holomorphic limit, zero. 
		Our starting point is the holomorphic anomaly equations and our goal is to take the $z\to 0$ limit of them. 
		If we denote by $D_z$ the covariant derivative on the moduli space (see \cite{Alim:2012gq} for more details) we have
		\begin{equation}
			\frac{\p F^{(0)_g}}{\p S^{zz}} = \frac{1}{2} D_z D_z F^{(0)}_{g-1} + \frac{1}{2} \sum_{h=1}^{g-1} D_z F^{(0)}_h \, D_z F^{(0)}_{g-h}.
		\end{equation}
		along with 
		\begin{equation}
			D_z S^{zz} = - C_{zzz}\,(S^{zz})^2 + h^{zz}_z, \quad
			\G^z_{zz}  = - C_{zzz}\,S^{zz} + s^z_{zz}, \quad
			D_z F^{(0)}_1 = \frac{1}{2} C_{zzz}\,S^{zz} + A^{(0)}_1.
			\label{eq:DzSzzSzzGammazzzSzzDzF01Szz}
		\end{equation}
		Here $h^{zz}_z$, $s^z_{zz}$ and $A^{(0)}_1$ are holomorphic ambiguities that can be fixed for each particular geometry. 
		We will assume that for small $z$ we can write
		\begin{equation}
			h^{zz}_z = h_0\,z + \CO(z^2), \qquad s^z_{zz} = \frac{s_0}{z} + \CO(z^0), \qquad A^{(0)}_1 = \frac{A_0}{z} + \CO(z^0).
		\end{equation}
		The Yukawa coupling is assumed to have the form
		\begin{equation}
			C_{zzz} = \frac{\k}{z^3\,\Delta(z)},
		\end{equation}
		where $\k$ is the classic intersection number and $\Delta(z)$ is a discriminant whose vanishing determines the conifold locus, satisfying $\Delta = 1 + \CO(z)$. 
		 
		In the holomorphic limit we have the relations
		\begin{equation}
			\left( \G^z_{zz} \right)_\hol = \left( \frac{\p T}{\p z}\right)^{-1} \left( \frac{\p^2 T}{\p z^2} \right) = - \frac{1}{z} + \CO(z^0),
		\end{equation}
		where $T \propto \log z + \CO(z^0)$ is the K\"ahler parameter, and 
		\begin{gather}
			\p_z \CF^{(0)}_1 = \p_z \log \left( \D^{-1/12}\,z^b (\det G_{z\overline{z}})^{-1/2} \right)= \frac{\tilde{b}}{z} + \CO(z^0) \\
			\tilde{b} := b + \frac{1}{2}, \qquad b = - \frac{1}{24} \int_{\text{CY}} c_2\wedge J.
			\label{eq:tildebb}
		\end{gather}
		Here $G_{z\overline{z}}$ is the Weil--Petersson metric on the complex structure moduli space with Levi-Civita connection $\G^z_{zz} = G^{z\overline{z}}\p_z G_{z\overline{z}}$. $J$ is the K\"ahler form and $c_2$ the second Chern class. 
		Taking the holomorphic limit of \eqref{eq:DzSzzSzzGammazzzSzzDzF01Szz} we find
		\begin{equation}
			h_0 = \k\,\s_\hol^2, \qquad s_0 = \k\,\s_\hol -1, \qquad A_0 = \tilde{b} - \frac{1}{2}\k\,\s_\hol.
			\label{eq:holo0relations}
		\end{equation}
		When $z\to 0$ the relevant nonholomorphic coordinate is $X$, satisfying $S^{zz} = z^2( \s_\hol + X)$. 
		We can calculate, using \eqref{eq:holo0relations}, 
		\begin{equation}
			\p_z X = \frac{1}{z}\,\k\,X^2 + \CO(z^0). 
		\end{equation}
		Taking into account that
		\begin{equation}
			\p_z F^{(0)}_g = \p_z X \, \p_X F^{(0)}_g + \CO(z^0), \qquad \p_{S^{zz}} F^{(0)}_g = \frac{1}{z^2}\p_X F^{(0)}_g,
		\end{equation}
		we arrive at the large radius limit of the holomorphic anomaly equations
		\begin{align}
			\p_X H^{(0)}_1 &= \frac{1}{2X}+\frac{\tilde{b}}{\k X^2}, 
			\label{eq:H01HAE} \\
			\p_X H^{(0)}_g &= \frac{3\k^2}{2} X^3 \p_X H^{(0)}_{g-1} + \frac{\k^2}{2} X^4 \p^2_X H^{(0)}_{g-1} + \frac{\k^2}{2} X^4 \sum_{h=1}^{g-1} \p_X H^{(0)}_{h}  \, \p_X H^{(0)}_{g-h}.
			\label{eq:H0gHAE}
		\end{align}
		The only explicit dependence on the CY geometry is through $\k$ and $\tilde{b}$. 
		This leads to the definition
		\begin{equation}
			X = \tilde{b}\k^{-1}u, \qquad H^{(0)}_g = \left( \tilde{b}^3 \k^{-1} \right)^{g-1} H^{(0),u}_g,
			\label{eq:XuHHu}
		\end{equation}
		so that the polynomials $H^{(0),u}_g(u)$ are now independent of the geometry.\footnote{We can still rescale $u$ and $H^{(0),u}_g$ by geometry-independent numbers. Here we make a particular choice such that the leading coefficients of $H^{(0),u}_g$ are the rational numbers $a^{(0)}_g$ relevant to \cite{Alim:2015qma}. Also, the integration constants arising from \eqref{eq:H0gHAE} must be zero to agree with $H^{(0),u}(u=0) = 0$. Recall that we subtracted the constant map contribution and $u=0$ represents the holomorphic limit in the large radius frame.} 

		Even though we have derived this result in the context of local geometries with a single complex structure modulus, the prescription is generic.\footnote{In the language of the nonholomorphic generators introduced in \cite{Alim:2013eja}, $T_2$ takes the role of $S^{zz}$. A similar definition for $u$ applies in that case with $T_2 =: T_{2,\hol} + \tilde{b}\k^{-1}u$. Note that no rescaling by $z^2$ is involved. } 
		See Table~\ref{tab:kappatildeballgeometries} for some examples.		
		\begin{table}[tb]
		\begin{center}
		\begin{tabular}{rcc}
			Geometry 	  & $\qquad\k\quad$ 			& $\quad\tilde{b}\quad$ \\[3pt]
			\hline \\[-7pt]
			local $\BP^2$   			    & $-\frac{1}{3}$  & $-\frac{1}{12}$  \\[3pt]
			local del Pezzo $E_5$ 		    & $-4$            & $\frac{1}{6}$    \\[3pt]
			local del Pezzo $E_6$ 		    & $-3$  		  & $\frac{1}{4}$    \\[3pt]
			local del Pezzo $E_7$ 		    & $-2$			  & $\frac{1}{3}$    \\[3pt]
			lcoal del Pezzo $E_8$ 		   	& $-1$ 			  & $\frac{5}{12}$   \\[3pt]
			mirror quintic  			    & $+5$			  & $-\frac{25}{12}$  \\[3pt]
			local $\BP^1\times\BP^1$	    & $-1$			  & $-\frac{1}{12}$
		\end{tabular}
		\end{center}
		\caption{Values of $\k$ and $\tilde{b}$ for local $\BP^2$, local del Pezzo geometries, the quintic mirror, and local $\BP^1\times\BP^1$. The last geometry has two moduli so the values of $\k$ and $\tilde{b}$ are effective (see Example~\ref{ex:localP1P1}).}
		\label{tab:kappatildeballgeometries}
		\end{table}
		\begin{example}[Local $\BP^1\times\BP^1$]\label{ex:localP1P1}
		In this geometry we have two moduli, $z_1$ and $z_2$ (see, for example, \cite{Haghighat:2008gw}).
		In the holomorphic limit the propagators tend to the same function, up to rescaling,
		\begin{equation}
			S^{11} \to S, \qquad S^{12} \to \frac{z_2}{z_1}\,S, \qquad S_{22} \to \frac{z_2^2}{z_1^2}\,S, \qquad S = z_1^2 \left( \frac{1}{2} - 2z_1 - 2z_2 + \ldots \right).
		\end{equation}
		If we consider the change of variables
		\begin{equation}
			S^{11} = z_1^2 \left( \frac{1}{2} + \frac{u}{12} \right), \qquad S^{12} = z_1 \,z_2 \left( \frac{1}{2} + \frac{u}{12} \right), \qquad S_{22} = z_2^2 \left( \frac{1}{2} + \frac{u}{12} \right) 
		\end{equation}
		we have that 
		\begin{equation}
			\lim_{z_1,z_2\to0} F^{(0),\BP^1\times\BP^1}_g = H^{(0),\BP^1\times\BP^1} = \left( 2^{-6}3^{-3} \right)^{g-1}\, H^{(0),u}_g.
		\end{equation}
		For geometries that do not effectively depend on an underlying genus 1 Riemann surface like this one we could have more than one variable $u$.
		\end{example}
		\begin{example}[Mirror quintic]
		For this compact geometry we must consider the rest of the propagators $S^z$ and $S$ \cite{Bershadsky:1993cx}, which acquire a nontrivial value in the holomorphic limit,
		\begin{equation}
			z^{-2} S^{zz}_\hol = -\frac{3}{25} + 173 z + \CO(z^2), \qquad z^{-1} S^z_\hol = \frac{2}{125} +  \CO(z), \qquad S_\hol = -\frac{3}{625} + \CO(z). 
		\end{equation}
		Note that we also have to rescale $S^z$ by $z$. 
		Letting
		\begin{equation}
			z^{-2}\,S^{zz} \equiv -\frac{3}{25} - \frac{5}{12}u, \qquad z^{-1}\,S^z \equiv \frac{2}{125}, 	\qquad S \equiv -\frac{3}{625}, 
		\end{equation}
		we find
		\begin{equation}
			\lim_{z\to0} F^{(0),\text{quintic}}_g = H^{(0),\text{quintic}}_g = \left(-5^5 2^{-6}3^{-3}  \right)^{g-1} H^{(0),u}_g.
		\end{equation}
		\end{example}

	\subsection{Transseries extension for rescaled free energies}
	\label{sec:transseriesextensiontorescaledfreeenergies}

		In this subsection we will obtain a transseries extension of the perturbative asymptotic series $H^{(0),u}(g_s;u)$. 
		We can take two complementary routes. 
		One was initiated in \cite{cesv13}, where the approach to the transseries extension of $F^{(0)}(g_s;z,S^{zz})$ was based on an extension of the holomorphic anomaly equations that admits transseries solutions.  
		The other route is based on \cite{Alim:2015qma} where a differential equation in $g_s$ was found and shown to be equivalent to the Airy equation. 
		The most important property of the transseries for $H^u$ is that the instanton corrections are not of the form $\exp(-A/g_s)$, but rather $\exp(-A/g_s^2)$. 
		This behavior is intimately related to the factorial growth $g!$ found in \cite{Couso-Santamaria:2014iia}, rather than $(2g)!$. 
		We will explore the implications of this later in Section~\ref{sec:closedstringinstantoncontributionstothetopologicalstring}. 

		The master equation for nonperturbative $F$ that was used in \cite{cesv13} is
		\begin{equation}
			\frac{\p F}{\p S^{zz}} + U \p_z F - \frac{1}{2}g_s^2 \left( D_z \p_z F + (\p_z F)^2 \right) = \frac{1}{g_s^2}W + V, 
			\label{eq:extendedHAEs}
		\end{equation}
		where $U$, $V$ and $W$ are determined by requiring the usual holomorphic anomaly equations to hold at the perturbative level (see \cite{cesv13} for details). 
		In the large radius limit this equation becomes
		\begin{equation}
			\p_u H^{u} - \frac{3}{2}g_s^2\,u^3 \left( \p_u H^{u} + \frac{u}{3} \p_u^2 H^{u} + \frac{u}{3} \left( \p_u H^{u} \right)^2 \right) = \frac{1}{2u} + \frac{1}{u^2}.
			\label{eq:uequation}
		\end{equation}
		We will refer to it as the $u$-equation. 
		The same equation can be obtained from \eqref{eq:H01HAE}-\eqref{eq:H0gHAE}, so $H^{(0),u}$ in \eqref{eq:uequation} actually includes the genus $1$ contribution. 

		It was found in \cite{Alim:2015qma} that the generating function 
		\begin{equation}
			\sum\limits_{g=2}^\infty \l_s^{2g-2} \,a^{(0)}_g
		\end{equation}
		satisfies a differential equation in $\l_s$ (c.f. Prop. 3.1). 
		Notice that the sum starts at genus $2$ instead of $1$. 
		A generalization of that equation can be obtained once we have understood the deformation in $u$ of $H^{(0),u}$. 
		This is done by taking a guess at the coefficients of $H^{(0),u}_g$. 
		If we denote
		\begin{equation}
			H^{(0),u} = \sum_{g=2}^\infty g_s^{2g-2}\,u^{2g-3}\sum_{p=0}^g a^{(0)}_{g,p}\,u^p,
		\end{equation}
		we can identify \footnote{If $f(\xi) = \sum f_n \, \xi^n$, then $[\xi^n](f(\xi)) := f_n$.}
		\begin{equation}
			a^{(0)}_{g,0} = \frac{1}{2^{g-2}}\frac{(2g-4)!}{g!(g-2)!}, \qquad a^{(0)}_{g,1} = \frac{2^{g-3}}{g-1}, \qquad a^{(0)}_{g,p} = a^{(0)}_p\, [\xi^g]\left( \frac{\xi^p}{(1-2\xi)^{\frac{3}{2}(p-1)}} \right).
		\end{equation}
		Note that the asymptotic nature of $H^{(0),u}$ comes from the coefficients $a^{(0)}_p$ alone. 
		Now we can follow \cite{Alim:2015qma} and obtain
		\begin{equation}
			\theta_{\tau_s}^2 H^{u} + \left( \theta_{\tau_s} H^{u}\right)^2 + \left( 1- \frac{2}{3\tau_s} - \frac{\tau_s}{u+\tau_s} \right) \theta_{\tau_s} H^{u} + \frac{5}{36} + \frac{1}{3u^2} + \frac{2\tau_s}{9u^3}+\frac{1}{6}\frac{2-\tau_s}{u+\tau_s} = 0.
			\label{eq:tausequation}
		\end{equation}
		where $\tau_s := g_s^2 u^3 = \l_s^2$, and $\theta_{\tau_s} = \tau_s \p_{\tau_s}$.
		The reason for introducing $\tau_s$ is to stress that the relevant (resurgent) variable is $g_s^2$ rather than $g_s$. 
		We will refer to \eqref{eq:tausequation} as the $\tau_s$-equation. 
		This equation itself cannot be cast into Airy form but we will go around this obstacle in Section~\ref{sec:largeordergrowthandresurgenceproperties}.  

		Next we introduce the transseries ansatz\footnote{In \eqref{eq:Hutransseries} the transseries parameter $\sigma$ is one of the two integration constants of the equation. The other has been implicitly fixed to reproduce the familiar perturbative series.} in $g_s^2$,
		\begin{equation}
			H^u(g_s^2;u) := H^{(0),u}(g_s^2;u) + \sum_{n=1}^\infty \sigma^n\,\rme^{-A_u(u)/g_s^2} \sum_{g=0}^\infty (g_s^2)^{g+b^{(n)}} \, H^{(n),u}_g(u).
			\label{eq:Hutransseries}
		\end{equation}
		and we solve for it order by order:
		
		\textsl{Perturbative:} The $u$-equation determines $H^{(0),u}_g(u)$ up to a holomorphic ambiguity, a constant, fixed by imposing that the free energy vanishes when $u=0$, since we have subtracted the constant map.  
		The $\tau_s$-equation already incorporates this condition at $u=0$ so no fixing is needed. 
		The form of the free energies is
		\begin{equation}
			H^{(0),u}_g(u) = u^{2g-3} \, \Pol(u;g),
		\end{equation}
		where the leading coefficient of the polynomial is $a^{(0)}_g$. 
		The first few energies were displayed in \eqref{eq:H0u2}-\eqref{eq:H0u6}. 

		\textsl{Instanton action:} The $u$-equation fixes the instanton action to be $A_u(u) = \frac{2}{3u^3} + \text{const}$, whereas the $\tau_s$-equation allows either $A_u(u) = \frac{2}{3u^3}$ or $A_u(u) = 0$.
		The only nontrivial solution is then
		\begin{equation}
			A_u(u) = \frac{2}{3u^3}.
		\end{equation}

		\textsl{One-instanton:} The $u$-equation for the one-instanton sector gives a tower of differential equations that determine $H^{(1),u}_g(u)$ up to a constant. 
		The $\tau_s$-equation does not determine the first coefficient $H^{(1),u}_0(u)$ at all, as is expected from this type of differential equations, but once this is fixed the rest follow. 
		At this point we choose to set $H^{(1),u}_0 = -\rme^{2/u}$ because it is compatible with the $u$-equation and with the resurgent properties we will describe in Section~\ref{sec:largeordergrowthandresurgenceproperties}. 
		Putting everything together we find
		\begin{equation}
			H^{(1),u}_g(u) = \rme^{2/u} \, u^g \, \Pol(u;2g). 
		\end{equation}
		The exponential term in front is determined by the $u$-equation. 
		The $\tau_s$-equation imposes $b^{(1)}=0$.
		The first few energies are
		\begin{align}
			H^{(1),u}_1 &= \rme^{2/u} \, u \, \left( \frac{5}{12}u^2 +1 \right), \\ 
			H^{(1),u}_2 &= \rme^{2/u} \, u^2 \, \left( -\frac{25 u^4}{288}+\frac{5 u^3}{4}-\frac{5 u^2}{12}+\frac{u}{3}-\frac{1}{2} \right).
		\end{align}

		\textsl{Higher-instanton:} Both the $u$-equation and $\tau_s$-equation for $H^{(n),u}_g$, $n \geq 1$, are algebraic so they involve no integration constants. 
		They give the same solution:
		\begin{equation}
			H^{(n),u}_g(u) = \rme^{n2/u} \, u^g \, \Pol(u;2g).
		\end{equation}			
		The $\tau_s$-equations impose $b^{(n)} = 0$. 
		The first few energies for the two-instanton sector are
		\begin{align}
			H^{(2),u}_0 &= -\rme^{4/u} \, \frac{1}{2}, \\
			H^{(2),u}_1 &= \rme^{4/u} \, u \, \left( \frac{5 u^2}{12} + 1 \right), \\ 
			H^{(2),u}_2 &= \rme^{4/u} \, u^2 \, \left( -\frac{25 u^4}{144}+\frac{5 u^3}{4}-\frac{5 u^2}{6}+\frac{u}{3}-1 \right).
		\end{align}

	\subsection{Large-order growth and resurgence properties}
	\label{sec:largeordergrowthandresurgenceproperties}

		We turn our attention to the resurgent properties of this transseries, that is, to the study how all the coefficients in \eqref{eq:Hutransseries} are related to each other. 
		To show some of the computational features of resurgence we will first \emph{extract} some coefficients encoded in the perturbative factorial growth, and only afterwards will we justify the results from a resurgent analysis of the Airy equation. 

		\subsubsection*{Large-order growth}
		
			We start off in the limit $u\to\infty$ and $\tau_s$ fixed that selects the coefficients $a^{(n)}_g$ from the transseries.
			One of the practical outcomes of resurgence is an explicit set of tight constraints between perturbative and nonperturbative coefficients of a transseries. 
			It is a relation of large order, meaning that the coefficients $a^{(m)}_h$, $h=0,1,2,\ldots$ of one sector are encoded in the growth of coefficients of another sector, $a^{(n)}_g$ when $g\to\infty$. 
			Usually we use the growth of perturbative coefficients, $n=0$, to find about nonperturbative coefficients, $m=1,2,\ldots$, but this type of large order relations hold between nonperturbative sectors as well. 
			In practice we do not have infinite coefficients but with enough of them we can obtain good numerical approximations. 
			The numerical procedure is based on Richardson extrapolation of sequences that converge to the number we wish to extract. 
			See \cite{Marino:2007te} for details of this method as applied in the context of matrix models. 
 
\begin{figure}[t]
				\centering
				\includegraphics[width=1.00\textwidth]{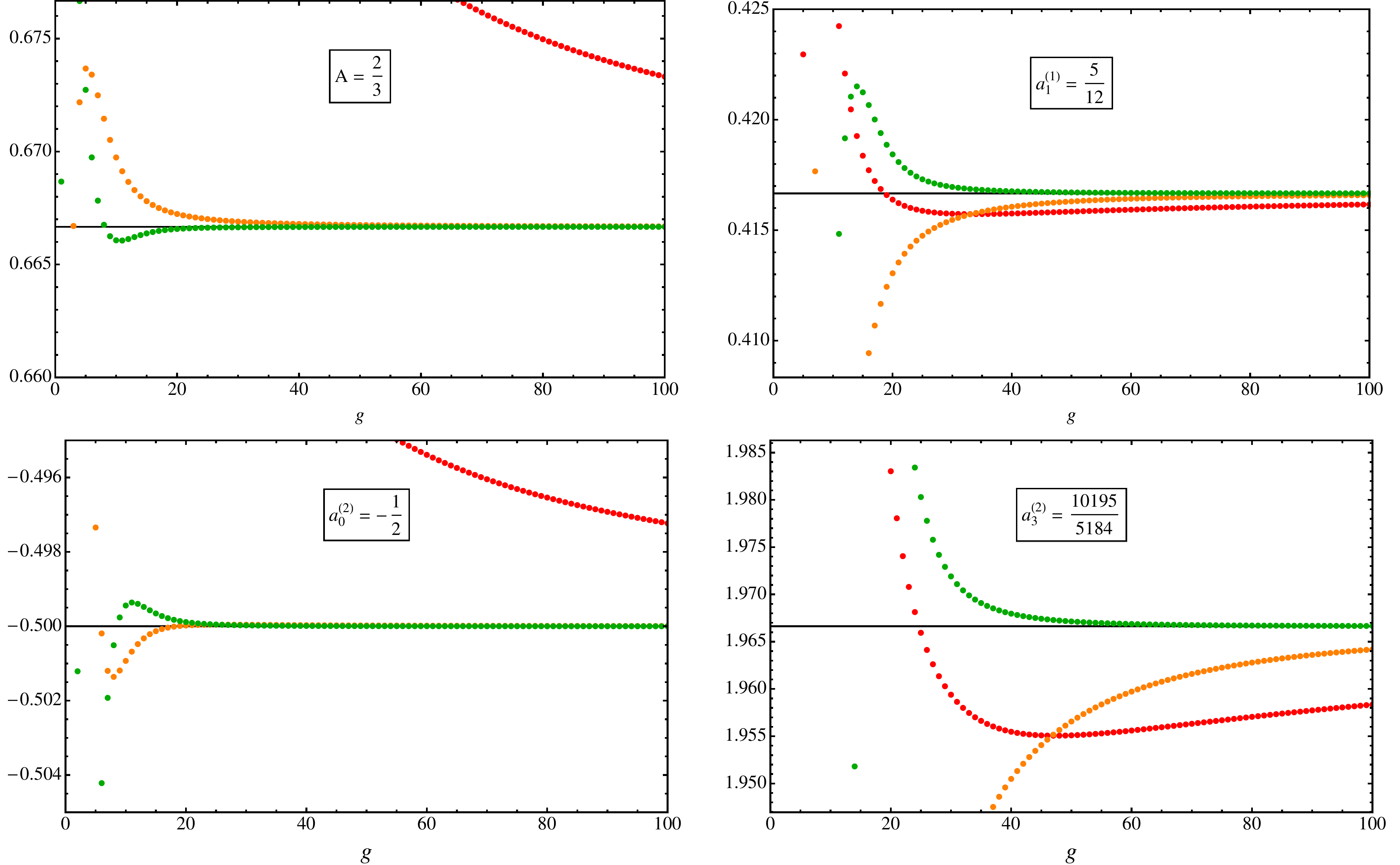}
				\caption{Graphical representation of large-order limits, $g\to\infty$, that converge to the instanton action and higher-instanton coefficients in \eqref{eq:a0gloexperimental} and \eqref{eq:a1gloexperimental}. 
				The original sequence (red) converging to the exact value is accelerated using Richardson extrapolation (orange and green). 
				}
				\label{fig:largeorderaplots}
			\end{figure}

			For example, by numerically analyzing how the numbers $a^{(0)}_g$ grow with $g$ --- see the top plots in Figure~\ref{fig:largeorderaplots} --- we experimentally find the relation
			\begin{equation}
				a^{(0)}_g \sim \frac{-\rmi}{2\pi\rmi}\left[\frac{\G(g-1)}{(2/3)^{g-1}}(-1) + \frac{\G(g-2)}{(2/3)^{g-2}}\left(\frac{5}{12}\right)+ \frac{\G(g-3)}{(2/3)^{g-3}}\left(-\frac{25}{288}\right) + \ldots \right],
				\label{eq:a0gloexperimental}
			\end{equation}
			but the numbers $-1$, $\frac{5}{12}$, $-\frac{25}{288}$, \ldots are none other than $a^{(0)}_h$, for $h=0,1,2,\ldots$, and $2/3$ is the instanton action of the transseries. 
			The number in front, $-\rmi$ in this case, is called the Stokes constant, $S_1$, and varies from problem to problem. 
			More compactly we write
			\begin{equation}
				a^{(0)}_g \sim \frac{S_1}{2\pi\rmi} \sum_{h=0}^\infty \frac{\G(g-1-h)}{A^{g-1-h}}\,a^{(1)}_h, \quad\text{as $g\to\infty$}.
			\end{equation}
			We can do something similar for the one-instanton coefficients, expecting to uncover the two-instanton sector:
			\begin{equation}
				a^{(1)}_g - \frac{1}{2\pi} \frac{\G(g)}{(-2/3)^g} \sim \frac{-\rmi}{\pi\rmi} \left[ \frac{\G(g)}{(2/3)^g} \left(-\frac{1}{2}\right) + \frac{\G(g-1)}{(2/3)^{g-1}} \left(\frac{5}{12}\right) + \frac{\G(g-2)}{(2/3)^{g-2}} \left(-\frac{25}{144}\right) + \ldots \right]
				\label{eq:a1gloexperimental}
			\end{equation}
			that we identify as
			\begin{equation}
				a^{(1)}_g \sim \frac{1}{2\pi} \frac{\G(g)}{(-A)^g} + \frac{S_1}{\pi\rmi} \sum_{h=0}^\infty \frac{\G(g-h)}{A^{g-h}}\,a^{(2)}_h, \quad\text{as $g\to\infty$}.
			\end{equation}
			See the bottom plots in Figure~\ref{fig:largeorderaplots} for some examples.  
			We will justify the extra term in the l.h.s. of \eqref{eq:a1gloexperimental} when we perform the resurgent analysis of $H^u$. 
	
\begin{figure}[t]
				\centering
				\includegraphics[width=1.00\textwidth]{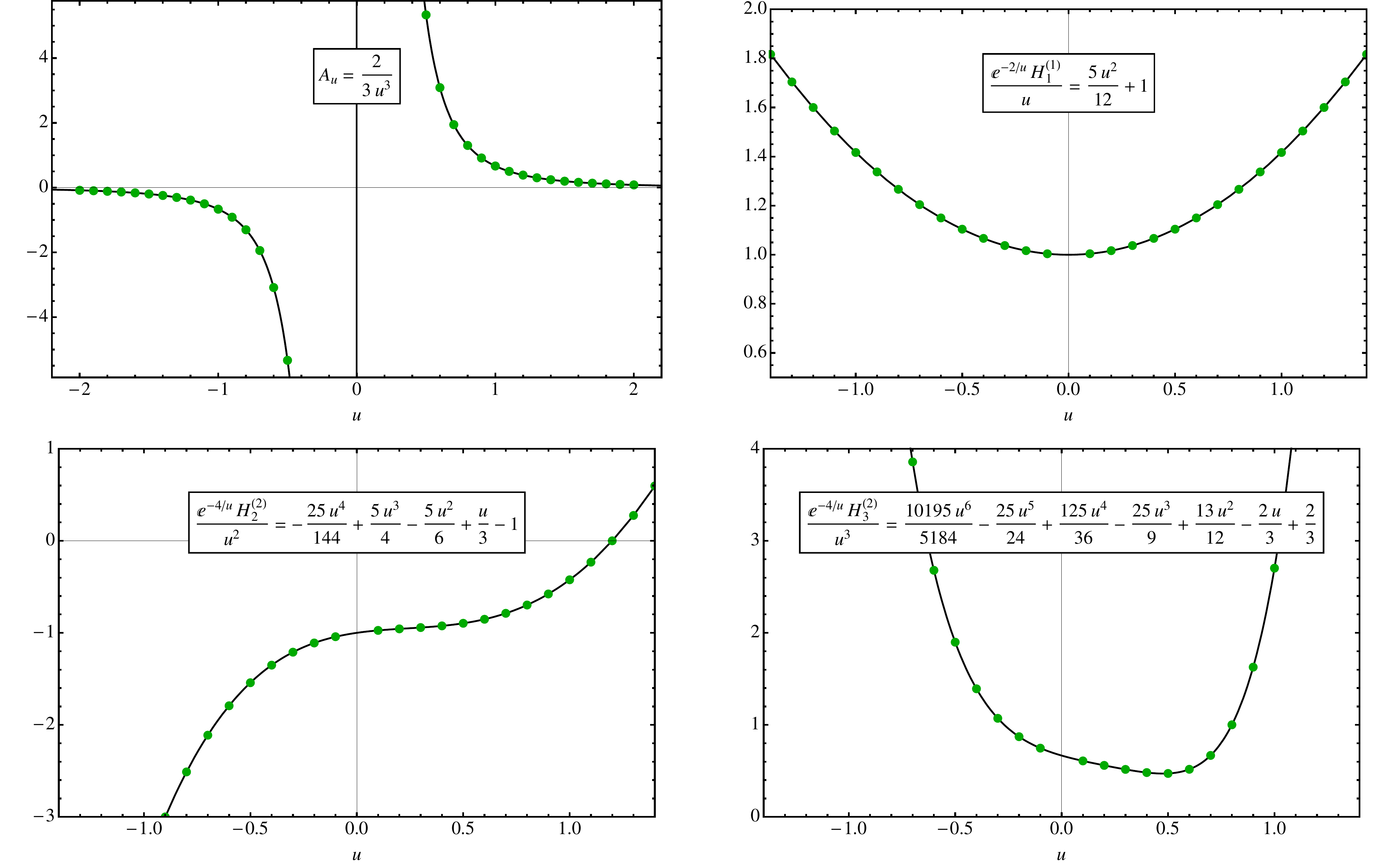}
				\caption{
				Graphical representation of large-order limits, $g\to\infty$, that converge to the instanton action and higher-instanton coefficients as functions of $u$.  
				Green dots come from the Richardson extrapolation of a sequence similar to the ones in Figure~\ref{fig:largeorderaplots}, whereas the solid lines are the transseries predictions described in Section~\ref{sec:transseriesextensiontorescaledfreeenergies}}
				\label{fig:largeorderHplots}
			\end{figure}

			We can take the same approach when $u$ is finite and find similar formulae
			\begin{align}
				H^{(0),u}_g &\sim \frac{S_1}{2\pi\rmi} \sum_{h=0}^\infty \frac{\G(g-1-h)}{A_u^{g-1-h}}\,H^{(1),u}_h, \quad\text{as $g\to\infty$},
				\label{eq:H0uglo} \\
				H^{(1),u}_g &\sim -\frac{1}{2\pi} \frac{\G(g)}{(-A_u)^g} + \frac{S_1}{\pi\rmi} \sum_{h=0}^\infty \frac{\G(g-h)}{A_u^{g-h}}\,H^{(2),u}_h, \quad\text{as $g\to\infty$}.
				\label{eq:H1uglo}
			\end{align}
			where $S_1 = -\rmi$ and $A_u = \frac{2}{3u^3}$. 
			See Figure~\ref{fig:largeorderHplots} for numerical verifications. 
			The large-order game does not end here, one could go on and uncover all coefficients of the transseries if computational power and memory were no obstacle. 
	
			\subsubsection*{Resurgence}
			There is a very nice justification for these large-order relations based on the properties of the Airy equation. 
			It was shown in \cite{Alim:2015qma} that the partition $\CZ^{(0)} = \exp \sum a^{(0)}_g \,\l_s^{2g-2}$ satisfies the Airy equation in the form
			\begin{equation}
				(\p_x^2 - x)v(x) = 0,
			\end{equation}
			where we have defined
			\begin{equation}
				x = \frac{1}{(2\l_s^2)^{2/3}}, \qquad v = 2^{-1/3}\,\rme^{\frac{1}{3\l_s^2}}\,\l_s^{1/3}\,\CZ^{(0)}. 
			\end{equation}
			When the parameter $u$ is present we can still obtain the same equation if we keep the quantity $\zeta := 1 - 2 \tau_s/u$ as a constant external parameter\footnote{The equation for $\tilde{H}^{(0),u}$ in that case is
			\begin{equation}
				\theta_{\tau_s}^2 \tilde{H}^{(0),u} + \left( \theta_{\tau_s} \tilde{H}^{(0),u}\right)^2 + \left( 1- \frac{\zeta - 1/3}{\tau_s} \right) \theta_{\tau_s} \tilde{H}^{(0),u} + \frac{5}{36} = \frac{1}{\l_s^4} \frac{(\zeta-1)^2(4\zeta-1)}{36},
			\end{equation}
			where the combination $\zeta = 1 - 2 \tau_s/u$ is kept fixed. 
			} and define
			\begin{equation}
				x = \frac{\zeta}{(2\tau_s)^{2/3}}, \qquad v = 2^{-1/3}\,\rme^{\frac{\zeta-1/3}{2\tau_s}}\,\tau_s^{1/6}\,\CZ^{(0),u},
				\label{eq:xvdefs}
			\end{equation}
			where $\CZ^{(0),u} := \exp \tilde{H}^{(0),u}$ and $\tau_s = \l_s^2$. 
			The tilde in $\tilde{H}^{(0),u}$ is to remind us that $\zeta$ is kept fixed. 
	
			The general transseries solution to the Airy equation is a combination of
			\begin{align}
				v_{\text{Bi/Ai}}(x) &= \frac{1}{2\sqrt{\pi}x^{1/4}}\rme^{\pm\frac{2}{3}x^{3/2}} \Phi_{\pm}(x^{3/2}), \\
				\intertext{where}
				\Phi_{\pm}(x) = \sum_{g=0}^\infty (\mp 1)^g c_g \, x^{-g},& \qquad c_g = \frac{1}{2\pi}\left( -\frac{3}{4} \right)^g \frac{\G(g+1/6)\G(g+5/6)}{g!}.
			\end{align}
			The perturbative solution $\CZ^{(0),u}$ is associated to $v_{\text{Bi}}$ \cite{Alim:2015qma}. 
			The nonperturbative one is 
			\begin{equation}
				\CZ^{u} = \sqrt{2\pi} \, 2^{1/3} \, \tau_s^{-1/6} \, \rme^{-\frac{\zeta-1/3}{2\tau_s}} \, \left( v_{\text{Bi}}\left( \frac{\zeta}{(2\tau_s)^{2/3}} \right) + \sigma v_{\text{Ai}}\left( \frac{\zeta}{(2\tau_s)^{2/3}} \right) \right),
				\label{eq:CZuvBiAi}
			\end{equation}
			as can be checked directly by expanding $v_{\text{Ai/Bi}}$ and comparing with the results in Section~\ref{sec:transseriesextensiontorescaledfreeenergies}. 
			In particular we can recover all the functions $H^{(n),u}_g(u)$ from the transseries expansion of $\log \CZ^{u}$ in $\tau_s$. 
			Since the prefactor in \eqref{eq:CZuvBiAi} has only trivial resurgent properties, the large-order relations found in this section must be explained by the Airy series alone. 
			If the goal is to arrive at relations \eqref{eq:H0uglo}-\eqref{eq:H1uglo} the start is to understand the singularities of the Borel transforms of $\Phi_\pm(x)$. 
			To connect the two we must introduce the concepts of alien derivative and Stokes automorphism. 
	
			The alien derivative, $\Delta_\omega$, on an asymptotic series captures the residues of its Borel transform at a singularity $\omega$. 
			For example, the Borel transform of $\Phi_\pm$ is
			\begin{equation}
				\CB[\Phi_\pm](s) = \sum_{g=1} \frac{(\mp1)^g c_g}{(g-1)!} s^{g-1} = \pm \frac{5}{48} {}_2F_1\left( \frac{7}{6}, \frac{11}{6}, 2; \pm\frac{3s}{4} \right),
			\end{equation}
			so we can compute the expansion around the singularity
			\begin{equation}
				\CB[\Phi_\pm]\left(s \pm \frac{4}{3} \right) = \frac{-\rmi}{2\pi\rmi s} - \rmi \CB[\Phi_\mp](s) \frac{\log(s)}{2\pi\rmi} + \text{regular}.
			\end{equation}
			The alien derivative on $\Phi_\pm$ collects\footnote{The actual definition for the alien derivative takes into account the choice of analytic continuation of the Borel transform around the singularities. In this case there is only one and the definition collapses to the one we are using.} the residues\footnote{Note that the Borel transform leaves out $c_0$ which reappears as the proper residue times $-\rmi$. } $-\rmi$ and $-\rmi \CB[\Phi_\mp]$ at $\pm\tilde{A} = \pm4/3$,
			\begin{equation}
				\Delta_{\pm \tilde{A}} \Phi_{\pm} = -\rmi \Phi_{\mp}.
			\end{equation}
			The factor $-\rmi$ is the Stokes constant. 
			The Airy system has quite simple resurgent properties because it comes from a linear differential equation. 
	
			Let us translate this result to $H$-language. 
			Using \eqref{eq:CZuvBiAi} we can write
			\begin{align}
				H^u &= \tilde{H}^{(0),u}(x;u) + \sum_{n=1}^\infty \sigma^n \, \rme^{-n \tilde{A} x^{3n/2}} \, \tilde{H}^{(n),u}(x;u) \nonumber\\
				&= \text{(non-resurgent)} + \log[\Phi_+(x)] + \sum_{n=1}^\infty \sigma^n \, \rme^{-n \tilde{A} x^{3n/2}} \frac{(-1)^{n+1}}{n} \left( \frac{\Phi_-}{\Phi_+} \right)^n.
				\label{eq:HuPhiMinusPhiPlus}
			\end{align}
			where the first term has a trivial alien derivative, so it can be ignored in what follows, and against the tilde in $\tilde{H}$ indicates that $\zeta = 1-2\tau_s/u$ is kept fixed. 
			The alien derivatives satisfy the Leibniz rule so we readily find
			\begin{align}
				\Delta_{+\tilde{A}} \tilde{H}^{(n),u} &= -\rmi\,(n+1) \, \tilde{H}^{(n+1),u}, \quad n\geq0, \\
				\Delta_{-\tilde{A}} \tilde{H}^{(0),u} &= 0, \\
				\Delta_{-\tilde{A}} \tilde{H}^{(1),u} &= -\rmi, 
				\label{eq:DeltaminusAH1u}\\
				\Delta_{-\tilde{A}} \tilde{H}^{(n),u} &= +\rmi\,(n-1) \, \tilde{H}^{(n-1),u}, \quad n\geq2.
			\end{align}
			To finally arrive at the large-order relations we use a dispersion relation argument going back to \cite{Simon:1970aa,Bender:1973rz}, that relates factorial growth and singular behavior 
			\begin{equation}
				\tilde{H}^{(n),u}(w) = \frac{1}{2\pi\rmi} \int_0^\infty dy \frac{\operatorname{Disc}_0\tilde{H}^{(n),u}(y)}{y-w} + \frac{1}{2\pi\rmi} \int_{-\infty}^0 dy \frac{\operatorname{Disc}_\pi \tilde{H}^{(n),u}(y)}{y-w}.
				\label{eq:dispersionrelation}
			\end{equation}
			Here we let $w := x^{3/2}$ to avoid writing fractional powers in the expressions.
			The discontinuities along the half-lines at angles $\theta=0,\pi$ can be computed in terms of alien derivatives through the Stokes automorphism, $\mathfrak{S}_\theta = 1 - \operatorname{Disc}_\theta$,
			\begin{equation}
				\mathfrak{S}_{0,\pi} = \exp \left( \rme^{\mp \tilde{A} w} \Delta_{\pm \tilde{A}} \right).
			\end{equation}
			Let us focus on the perturbative ($n=0$) and one-instanton ($n=1$) sectors. 
			We immediately calculate
			\begin{align}
				\mathfrak{S}_0 \tilde{H}^{(0),u} &= \tilde{H}^{(0),u} + \sum_{n=1}^\infty \rme^{-n \tilde{A} w} (-\rmi)^n \tilde{H}^{(n),u}, 
				\label{eq:frakS0H0u}\\
				\mathfrak{S}_\pi \tilde{H}^{(0),u} &= \tilde{H}^{(0),u}, \\
				\mathfrak{S}_0 \tilde{H}^{(1),u} &= \tilde{H}^{(1),u} + \sum_{n=1}^\infty \rme^{-n \tilde{A} w} (-\rmi)^n(n+1)\tilde{H}^{(n+1),u}, \\
				\mathfrak{S}_\pi \tilde{H}^{(1),u} &= \tilde{H}^{(1),u} - \rme^{+\tilde{A} w} \, \rmi.
				\label{eq:frakSpiH1u}
			\end{align}
			These equations tell us that the large-order growth of $\tilde{H}^{(0),u}_g$ is only affected by the pole at $\tilde{A}$, receiving contributions from all instanton orders, while $\tilde{H}^{(1),u}_g$ is also influenced by the pole at $-\tilde{A}$ due to \eqref{eq:DeltaminusAH1u}. 
			This type of inhomogeneous term is not found in resurgent systems such as matrix models. 
			In this case it arises because of the structure of $H^u$ shown in \eqref{eq:HuPhiMinusPhiPlus} involving the ratio $\Phi_-/\Phi_+$. 
			Similar inhomogeneous terms will also appear in the equivalent expressions of \eqref{eq:frakSpiH1u} for higher instanton sectors, $n\geq 2$. 
			This happens because the exponential expansion of $\mathfrak{S}_\pi$ on $H^{(n),u}$ truncates after $n+1$ terms and the last one is proportional to \eqref{eq:DeltaminusAH1u}. 

			If we now expand \eqref{eq:dispersionrelation} around large $x$ (there, $w = x^{3/2}$) and use the equations \eqref{eq:frakS0H0u}-\eqref{eq:frakSpiH1u} we arrive at the large-order relations described earlier for $a^{(n)}_g$ when $\zeta = 1$ and for any $u$.\footnote{The instanton action $\tilde{A} = 4/3$ becomes $A = 2/3$ once we take the factor of $2$ in \eqref{eq:xvdefs} between $x$ and $\tau_s$. Removing the constraint that $\zeta$ is fixed to go from $\tilde{H}$ to $H$ takes some extra work. Alternatively one can take a resurgence approach to \eqref{eq:tausequation}; see also the comments below. } 

		\subsubsection*{Comments}

			All the resurgent properties of the rescaled $u$-dependent free energies stem directly from those of the Airy system. 
			A complete resurgent analysis like we have presented here is not available from the alternative viewpoint of the holomorphic anomaly equation. 
			That was the only approach at hand in \cite{Couso-Santamaria:2014iia} so the analysis was necessarily limited to the results obtained from a large-order analysis, leaving open questions regarding the ultimate resurgent structure of the topological string free energy $F(z,S^{zz})$. 
	
			In this way the opportunity of having two equations, one in the string coupling and one in the antiholomorphic moduli, could provide some insight to the problem just mentioned.
			A well adapted technique might be parametric resurgence \cite{e84} (also dubbed coequational or quantum). 
			As the name suggests, it deals with resurgent systems in which the resurgent variable appears as a parameter. 
			This is exactly the case for the holomorphic anomaly equation although we will not pursue this path further in this work.
	
			Finally, let us stress the fact that the transseries considered here and in \cite{cesv13} are different, one involving $\rme^{-1/g_s^2}$ and the other $\rme^{-1/g_s}$. 
			From a wider perspective these are two different sectors of the same problem. 
			We go deeper into this matter in the next section.

\section{Instanton contributions of NS-brane type}
\label{sec:closedstringinstantoncontributionstothetopologicalstring}

	We have found in the previous section that the rescaled perturbative free energies, $H^{(0),u}_g$, have a milder factorial growth with $g$ than the original, moduli dependent free energies, $F^{(0)}_g$, namely $g!$ as opposed to $(2g)!$. 
	The natural transseries extensions that dictate these particular growths are of different type. 
	The growth of $F^{(0)}_g$ comes with a transseries extension of the form \eqref{eq:Fusualtransseries} and instanton corrections of the form $\rme^{-A_\text{D}/g_s}$. 
	We label these nonperturbative effects of D-brane type because such a dependence in $g_s$ is the natural one for D-branes \cite{Polchinski:1994fq}. 
	In some theories it can be checked that D-branes are responsible for these corrections, see for example \cite{Martinec:2003ka,Alexandrov:2003nn,Marino:2006hs,Marino:2007te,Pasquetti:2009jg}. 
	By contrast, the transseries that completes the rescaled free energies, equation \eqref{eq:Hutransseries}, has corrections of the form $\rme^{-A_u/g_s^2}$, which are natural in the context of NS-branes in string theory \cite{Strominger:1990et,Callan:1991dj,Callan:1991ky}. 
	We want to explore if there exists a moduli dependent transseries with the same $g_s$ dependence as that of $H^u$, 
	\begin{equation}
		F = \sum_{g=0}^\infty g_s^{2g-2} F^{(0)}_g(z,\overline{z}) + \sigma_\text{NS} \, \rme^{-A_\text{NS}(z,\overline{z})/g_s^2} \sum_{g=0}^\infty (g_s^2)^{g} F^{(1)}_g(z,\overline{z}) + \cdots
		\label{eq:closedstringtransseriesansatz}
	\end{equation}
	that could complement the one studied in \cite{cesv13} and shown in  \eqref{eq:Fusualtransseries}. 
	That is, we want to answer the question: does the transseries extension of the perturbative string free energy include sectors of NS-brane type like those in \eqref{eq:closedstringtransseriesansatz}?
	
	There are at least two ways to give a positive answer to this question. 
	The first would be to find signs of $g!$-growth for $F^{(0)}_g$ outside the large radius limit. 
	This requires resumming the leading $(2g)!$ contributions from terms weighed by $\rme^{-nA/g_s}$, for all $n$, so it looks impractical. 
	The second option would look at the resummation of the free energy transseries, $F$ \cite{Couso-Santamaria:2015aa}. 
	If the instanton contributions of NS-brane type are present they may contribute to the resummation and they could be detected this way. 
	Note that even if these new sectors are present in the formal transseries, this does not mean that they will be present in the resummation because $\sigma_\text{NS}$ could be zero.\footnote{Then again, Stokes phenomenon has the generic property of turning on the value of the transseries parameter $\sigma_\text{NS}$ so there may be values of $z$ and $S^{zz}$ for which these sectors are visible through resummation. }
	A different approach would provide physical arguments for such sectors based on the existence of instanton-like objects in topological string theory that could account for them. 
	We will speculate at the end about this possibility having in mind NS5-branes as the possible origin of these effects. 
	Let us first explore some of the consequences of the answer to the question of existence of NS-brane effects being affirmative. 

	\subsubsection*{Transseries of NS-brane type}
	
		The master equation \eqref{eq:extendedHAEs}, explored in \cite{cesv13} to study transseries with open string instanton sectors, actually admits transseries solutions of the form \eqref{eq:closedstringtransseriesansatz}. 
		The equation for the instanton action $A_\text{NS}$ is
		\begin{equation}
			\p_{S^{zz}} A_\text{NS} + \frac{1}{2} \left( \p_z A_\text{NS} \right)^2 = 0. 
			\label{eq:eqforAcl}
		\end{equation}
		This contrasts with the holomorphicity condition found in \cite{cesv13} for the instanton action $A_\text{D}$ where the second term in \eqref{eq:eqforAcl} was absent. 
		The equation for $A_\text{NS}$ is nonlinear and more difficult to solve. 
		If we look for a solution that matches $A_u$ in the large radius limit we can take as an ansatz
		\begin{equation}
			A_\text{NS} = \sum_{p=0}^\infty A_p(X) \, z^p, \qquad \text{with} \qquad A_0(X) = \frac{2}{3\k^2 X^3}.
			\label{eq:AclwithA0X}
		\end{equation}
		$A_0(X)$ is in correspondence with $A_u$ once we divide by $g_s^2$ and take \eqref{eq:XuHHu} into account. 
		The corrections in $z$, $A_p(X)$, are not universal and include new integration constants.  
		For example, 
		\begin{equation}
			A_1(X) = \text{const}_1\cdot\rme^{\frac{2}{\k X}} + \sum_{p=-4}^0 b_k \, X^k,
		\end{equation}
		where the numbers $b_k$ depend on the particular geometry. 
		To fix the constants we look at a different expansion for $A_\text{NS}$, in $(S^{zz})^{-1}$,
		\begin{equation}
			A_\text{NS} = \sum_{n=3}^\infty \frac{A_n(z)}{(S^{zz})^n}.
			\label{eq:ANSseriesSzz}
		\end{equation}
		In the example of local $\BP^2$ the solution reads 
		\begin{equation}
			A^{\BP^2}_n(z) = z^{2n}\,(1+27z)^2 \, \Pol\left(z;\left[\frac{2n}{3}\right]-2 \right). 
		\end{equation}
		Comparing with \eqref{eq:AclwithA0X} imposes ${\text{const}_1 = -3}$, and similarly with higher orders and other geometries. 
		The prefactor $(1+27z)^2$ can be identified with the squared discriminant, $\Delta(z)^2$, which suggests that in general this instanton action vanishes at the conifold locus.
		
		It is not clear that there are no other interesting solutions to \eqref{eq:eqforAcl}. 
		Note that a constant action is always a solution but a nontrivial holomorphic function, without propagator dependence, is not. 
		
		Finally, if we take \eqref{eq:AclwithA0X} we can solve for $F^{(1)}_g$, and find that their large radius limit is in correspondence with $H^{(1),u}_g$. 
		For example, since $H^{(1),u}_0 \propto \exp(2/u)$ we look for 
		\begin{equation}
			F^{(1)}_0 = \text{const}\cdot \rme^{G^{(1)}_0}, \qquad G^{(1)}_0 = \sum_{n=1}^\infty \frac{G^{(1)}_{0,n}(z)}{(S^{zz})^n}
		\end{equation}
		and solve algebraically for $G^{(1)}_{0,n}(z)$, finding polynomials in $z$. 
		When $z\to0$ we recover $H^{(1),u}_0$. 
	
	\subsubsection*{Seeking an interpretation for $A_\text{NS}$} 
		
		The holomorphic anomaly equation for $A_\text{NS}$, associated to instanton corrections of NS-brane type, has a solution with a well-defined rescaled limit, $A_u$, which is universal and is linked to the existence of a large radius point in the complex structure moduli space. 
		Even though we only have direct evidence for $A_u$ it is natural to think of it as the large radius limit of an action defined everywhere in moduli space. 
		So let us assume that such an instanton action $A_\text{NS}$ is a well-defined quantity and look for a physical interpretation for it. 

		For the case of NS5-branes the action $A_\text{NS}$ is proportional to the volume of the Calabi--Yau manifold wrapped by the brane \cite{Becker:1995kb}. 
		To check whether the volume is a solution of the equation for $A_\text{NS}$ let us rewrite \eqref{eq:eqforAcl} in $(z,\overline{z})$ coordinates,
		\begin{equation}
			\p_{\overline{z}} A_\text{NS} + \frac{1}{2} \overline{C}^{zz}_{\overline{z}} \, (\p_z A_\text{NS})^2 = 0,
		\end{equation}
		where $\overline{C}^{zz}_{\overline{z}} = \rme^{2K}\,(G^{z\overline{z}})^2\,\overline{C_{zzz}}$. 
		Recall that the metric in moduli space comes from a K\"ahler potential, that is $G_{z\overline{z}} = \p_z\p_{\overline{z}}K$. 
		The volume of the A-model Calabi--Yau manifold is given by the classical part of $\frac{1}{8}\rme^{-K}$, where by classical part we mean that we only keep perturbative terms in the K\"ahler parameter $T$ and discard terms in $\rme^{T}$ \cite{Candelas:1990rm,Strominger:1990pd}, 
		 \begin{equation}
		 	\frac{1}{8}\rme^{-K}  = \text{Vol} +  \CO\left( \rme^T, \rme^{\overline{T}} \right). 
		 \end{equation}
		The K\"ahler potential can be written in terms of periods of the nowhere vanishing {$(3,0)$-form} $\Omega$, as
		\begin{equation}
			\rme^{-K} \equiv \rmi W =\rmi \int_{\text{CY}} \Omega \wedge \overline{\Omega} = \rmi \left[ 2(F_0 - \overline{F_0}) - (T - \overline{T})(\p_T F_0 + \overline{\p_T F_0}) \right].
		\end{equation}
		So we can write the equation for $A_\text{NS}$ as 
		\begin{equation}
			(W \,\p_T\p_{\overline{T}}W - \p_T W\,\p_{\overline{T}}W )^2 \, \p_{\overline{T}} A_\text{NS} - \frac{1}{2} \overline{C_{TTT}}\,W^2\,(\p_T A_\text{NS})^2 = 0. 
			\label{eq:eqforANSW}
		\end{equation}
		The holomorphic prepotential $F_0$, that is, the genus-0 free energy, has the general form
		\begin{equation}
			F_0 = \frac{\kappa}{3!}T^3 + c_2 T^2 + c_1 T + \CO\left( \rme^T \right). 
			\label{eq:genericformF0}
		\end{equation}
		The numbers $c_2$, $c_3$ are irrelevant to what follows. 
		The nonclassical terms are subleading when we approach the large radius point. 
		From equation \eqref{eq:genericformF0} we can calculate
		\begin{equation}
			W = - \frac{\kappa}{3!}(T - \overline{T})^3 + \CO\left(\rme^T,\rme^{\overline{T}}\right)
		\end{equation}
		By direct computation we can see that $W$ is a solution of \eqref{eq:eqforANSW} up to nonclassical terms, that is
		\begin{equation}
			(W \,\p_T\p_{\overline{T}}W - \p_T W\,\p_{\overline{T}}W )^2 \, \p_{\overline{T}} W - \frac{1}{2} \overline{C_{TTT}}\,W^2\,(\p_T W)^2 = \CO\left(\rme^T,\rme^{\overline{T}}\right). 
		\end{equation}
		So we have found that a multiple of the Calabi--Yau volume is an approximate solution of the equation for the instanton action, in the limit where we approach the large radius point. 
		This approximate solution can be identified with the one described in \eqref{eq:ANSseriesSzz} by noticing that the antiholomorphic dependence\footnote{For example, in local $\BP^2$ the propagator in essentially proportional to $\hat{E}_2(\tau,\overline{\tau}) = E_2(\tau) - \frac{6\rmi}{\pi(\tau-\overline{\tau})}$ and $\tau = \p_T^2 F_0$ is proportional to $T$ to leading order.} of the propagator is roughly of the form
		\begin{equation}
			z^{-2}S^{zz} \propto \frac{1}{T-\overline{T}} + \CO\left( \rme^T \right). 
		\end{equation}

		The classical part\footnote{Both $W$ and its classical truncation are only approximate solutions the equation for the instanton action.} of $W = \int \Omega \wedge \overline{\Omega}$ is equal to the instanton action $A_\text{NS}$ when we approach the large-radius limit, $z=0$. 
		Note that the proportionality factor between $V$ and the classical part of $W$i includes an imaginary unit, $\rmi$, because $W$ is purely imaginary whereas the volume is real. 
		Even though there is a connection between $A_\text{NS}$ and the Calabi--Yau volume and this points towards an NS5-brane interpretation of the action, we must understand what are the corrections in $\rme^T$ and $\rme^{\overline{T}}$ that would satisfy the equation to all orders, and the meaning of the proportionality factor. 

		Let us finish by noticing that NS5-brane effects where found in \cite{Pioline:2009ia} manifested as the nonperturbative ambiguity in the resummation of D-brane instanton effects in string theory. 
		To leading order in $g_s$ these contributions can be regarded as a series in $\rme^{-A_\text{D}/g_s}$ where we sum over instanton sectors (or Ramond-Ramond charges in the case of string scattering amplitudes on Calabi--Yau compactifications). 
		This series is asymptotic, with Gaussian rather than factorial growth, and its nonperturbative ambiguity can be approximated by optimal truncation giving an effect $\rme^{-A_\text{NS}/g_s^2}$, where $A_\text{NS}$ was shown to be proportional to the Calabi--Yau volume. 
		It turns out that a similar calculation can be done using the terms with $g=0$ in the D-brane type of transseries \eqref{eq:Fusualtransseries}, namely
		\begin{equation}
			\sum_{n=1}^\infty F^{(n)}_0 \rme^{-n A_\text{D}/g_s}.
		\end{equation}
		It was found in \cite{cesv13} that $F^{(n)}_0 \propto \rme^{n^2\frac{1}{2}(\p_z A_\text{D})^2 S^{zz}}$, so the growth is also Gaussian in $n$. 
		This leads to a potential ambiguity of size
		\begin{equation}
			\exp\left(-\frac{1}{g_s^2} \frac{A_\text{D}^2}{2(\p_z A_\text{D})^2 S^{zz}} \right). 
		\end{equation}
		Note that the power of the propagator is not cubic so there is no obvious relation the Calabi--Yau volume. 
		This adds to the list of unanswered questions about the interpretation of $A_\text{NS}$ as a nonperturbative effect generated by actual NS-branes.

\section{Conclusions}

	We have seen that there is a well-defined large radius limit of the perturbative free energy for topological strings once the propagator is rescaled appropriately. 
	The rescaled free energies keep some antiholomorphic dependence and their functional form is universal, that is, independent of the Calabi--Yau geometry. 
	This limit includes the one considered in \cite{Alim:2015qma} and provides a geometric interpretation for the rescaling by associating it with the large radius point. 

	The rescaled free energies, $H^{(0),u}_g$, form an asymptotic series that grows like $g!$, as was found in \cite{Couso-Santamaria:2014iia}, in contrast with the $(2g)!$-growth typical of a topological expansion. 
	The fact that $\exp H^{(0),u}$ (times a simple function of $g_s$ and $u$) satisfies the Airy equation \cite{Alim:2015qma} can be used to explain the resurgent properties of $H^{(0),u}$ and its transseries extension. 
	An alternative approach to the latter comes from the large radius limit of the holomorphic anomaly equations. 

	The transseries for $H^u$ depends on $\rme^{-A_u/g_s^2}$ rather than $\rme^{-A_\text{D}/g_s}$. 
	For general values of the complex structure, away from the large radius point, the holomorphic anomaly equations admit a transseries solution with $\rme^{-A_\text{NS}/g_s^2}$, where $A_\text{NS}$ is not holomorphic. 
	There is a connection between $A_\text{NS}$ and the volume of the Calabi--Yau manifold in the large-radius limit which points to an NS5-brane interpretation although this relation must be understood better.

\section*{Acknowledgements}

	I would like to thank David Sauzin, Emanuel Scheidegger, Ricardo Schiappa, and Marcel Vonk for helpful discussions and comments. 
	I also appreciate useful comments and observations by Marcos Mari\~no, Boris Pioline, Ricardo Schiappa, and Marcel Vonk on a draft of this paper.
	This research is supported by the FCT-Portugal grant EXCL/MAT-GEO/0222/2012.

\bibliographystyle{myutphys}
\bibliography{paper}

\end{document}

%% file: newcommands.tex
\newcommand{\CB}{\mathcal{B}}

\newcommand{\CF}{\mathcal{F}}

\newcommand{\CO}{\ensuremath{\mathcal{O}}\xspace}

\newcommand{\CZ}{\mathcal{Z}}

\newcommand{\BP}{\mathbb{P}}

\newcommand{\be}{\begin{equation}}
\newcommand{\ee}{\end{equation}}
\newcommand{\bea}{\begin{eqnarray}}
\newcommand{\eea}{\end{eqnarray}}

\newcommand{\p}{\partial}

\newcommand{\rme}{{\mathrm{e}}}
\newcommand{\rmi}{{i\mkern1mu}}

\newcommand{\s}{\sigma}
\newcommand{\D}{\Delta}

\renewcommand{\l}{\lambda}

\theoremstyle{remark} \newtheorem{example}{Example}

\newtheorem{thm}{Theorem}

\newtheorem{cor}{Corollary}

\newcommand{\bthm}{\begin{thm}}
\newcommand{\ethm}{\end{thm}}
\newcommand{\bprop}{\begin{prop}}
\newcommand{\eprop}{\end{prop}}
\newcommand{\bcor}{\begin{cor}}
\newcommand{\ecor}{\end{cor}}
\newcommand{\bdefn}{\begin{defn}}
\newcommand{\edefn}{\end{defn}}
\newcommand{\brem}{\begin{rem}}
\newcommand{\erem}{\end{rem}}

\newcommand{\G}{\ensuremath{\Gamma}\xspace}

\def\hol{{\textrm{hol}}}

\def\1{\mathbf{1}}

\def\XXint#1#2#3{{\setbox0=\hbox{$#1{#2#3}{\int}$}
     \vcenter{\hbox{$#2#3$}}\kern-.5\wd0}}

\newcommand{\Pol}{{\text{Pol}}}

\renewcommand{\k}{{\kappa}}

\DeclarePairedDelimiter\abs{\lvert}{\rvert}%
\makeatletter
\let\oldabs\abs
\def\abs{\@ifstar{\oldabs}{\oldabs*}}

%% file: paper.bbl
\providecommand{\href}[2]{#2}\begingroup\begin{thebibliography}{10}

\bibitem{Gross:1988ib}
D.~J. Gross and V.~Periwal, ``String perturbation theory diverges,''
\href{http://dx.doi.org/10.1103/PhysRevLett.60.2105}{{\em Phys.Rev.Lett.}
  {\bfseries 60} (1988) 2105}.

\bibitem{shenker1990}
S.~Shenker, \href{http://dx.doi.org/10.1007/978-1-4615-3772-4_12}{``The
  strength of nonperturbative effects in string theory,''} in {\em Random
  surfaces and quantum gravity}, O.~Alvarez, E.~Marinari, and P.~Windey, eds.,
  vol.~262 of {\em NATO ASI Series}, pp.~191--200.
\newblock Springer US, 1991.

\bibitem{Bershadsky:1993ta}
M.~Bershadsky, S.~Cecotti, H.~Ooguri, and C.~Vafa, ``Holomorphic anomalies in
  topological field theories,''
  \href{http://dx.doi.org/10.1016/0550-3213(93)90548-4}{{\em Nucl.Phys.}
  {\bfseries B405} (1993) 279--304},
\href{http://arxiv.org/abs/hep-th/9302103}{{\ttfamily arXiv:hep-th/9302103
  [hep-th]}}.

\bibitem{Bershadsky:1993cx}
M.~Bershadsky, S.~Cecotti, H.~Ooguri, and C.~Vafa, ``{Kodaira--Spencer theory
  of gravity and exact results for quantum string amplitudes},''
  \href{http://dx.doi.org/10.1007/BF02099774}{{\em Commun.Math.Phys.}
  {\bfseries 165} (1994) 311--428},
\href{http://arxiv.org/abs/hep-th/9309140}{{\ttfamily arXiv:hep-th/9309140
  [hep-th]}}.

\bibitem{Yamaguchi:2004bt}
S.~Yamaguchi and S.-T. Yau, ``{Topological string partition functions as
  polynomials},'' \href{http://dx.doi.org/10.1088/1126-6708/2004/07/047}{{\em
  JHEP} {\bfseries 0407} (2004) 047},
\href{http://arxiv.org/abs/hep-th/0406078}{{\ttfamily arXiv:hep-th/0406078
  [hep-th]}}.

\bibitem{Alim:2007qj}
M.~Alim and J.~D. Lange, ``{Polynomial structure of the (open) topological
  string partition function},''
  \href{http://dx.doi.org/10.1088/1126-6708/2007/10/045}{{\em JHEP} {\bfseries
  0710} (2007) 045},
\href{http://arxiv.org/abs/0708.2886}{{\ttfamily arXiv:0708.2886 [hep-th]}}.

\bibitem{Alim:2012gq}
M.~Alim, ``{Lectures on mirror symmetry and topological string theory},''
\href{http://arxiv.org/abs/1207.0496}{{\ttfamily arXiv:1207.0496 [hep-th]}}.

\bibitem{Grassi:2014zfa}
A.~Grassi, Y.~Hatsuda, and M.~Mari\~no, ``{Topological strings from quantum
  mechanics},''
\href{http://arxiv.org/abs/1410.3382}{{\ttfamily arXiv:1410.3382 [hep-th]}}.

\bibitem{Marino:2015nla}
M.~Mari\~no, ``{Spectral theory and mirror symmetry},''
\href{http://arxiv.org/abs/1506.07757}{{\ttfamily arXiv:1506.07757 [math-ph]}}.

\bibitem{Ecalle:1981uy}
J.~\'Ecalle. ``Les fonctions resurgentes,'' \textsl{Pr\'epub. Math.
  Universit\'e} \textbf{81-05} (1981), \textbf{81-06} (1981), \textbf{85-05}
  (1985).

\bibitem{sauzin:2014aa}
D.~Sauzin, ``{Introduction to 1-summability and the resurgence theory},''
  \href{http://arxiv.org/abs/1405.0356}{{\ttfamily arXiv:1405.0356 [hep-th]}}.

\bibitem{Marino:2012zq}
M.~Mari\~no, ``{Lectures on non-perturbative effects in large $N$ gauge
  theories, matrix models and strings},''
  \href{http://dx.doi.org/10.1002/prop.201400005}{{\em Fortsch.Phys.}
  {\bfseries 62} (2014) 455--540},
\href{http://arxiv.org/abs/1206.6272}{{\ttfamily arXiv:1206.6272 [hep-th]}}.

\bibitem{cesv13}
R.~Couso-Santamar\'ia, J.~D. Edelstein, R.~Schiappa, and M.~Vonk, ``Resurgent
  transseries and the holomorphic anomaly,''
  \href{http://dx.doi.org/10.1007/s00023-015-0407-z}{{\em Annales Henri
  Poincar\'e} (2015) 1--69, in press},
\href{http://arxiv.org/abs/1308.1695}{{\ttfamily arXiv:1308.1695 [hep-th]}}.

\bibitem{Couso-Santamaria:2014iia}
R.~Couso-Santamar\'ia, J.~D. Edelstein, R.~Schiappa, and M.~Vonk, ``{Resurgent
  transseries and the holomorphic anomaly: nonperturbative closed strings in
  local ${\mathbb{C}\mathbb{P}^2}$},''
  \href{http://dx.doi.org/10.1007/s00220-015-2358-0}{{\em Commun.Math.Phys.}
  {\bfseries 338} no.~1, (2015) 285--346},
\href{http://arxiv.org/abs/1407.4821}{{\ttfamily arXiv:1407.4821 [hep-th]}}.

\bibitem{Alim:2015qma}
M.~Alim, S.-T. Yau, and J.~Zhou, ``{Airy equation for the topological string
  partition function in a scaling limit},''
\href{http://arxiv.org/abs/1506.01375}{{\ttfamily arXiv:1506.01375 [hep-th]}}.

\bibitem{Drukker:2011zy}
N.~Drukker, M.~Mari\~no, and P.~Putrov, ``{Nonperturbative aspects of ABJM
  theory},'' \href{http://dx.doi.org/10.1007/JHEP11(2011)141}{{\em JHEP}
  {\bfseries 1111} (2011) 141},
\href{http://arxiv.org/abs/1103.4844}{{\ttfamily arXiv:1103.4844 [hep-th]}}.

\bibitem{Alim:2013eja}
M.~Alim, E.~Scheidegger, S.-T. Yau, and J.~Zhou, ``{Special polynomial rings,
  quasi modular forms and duality of topological strings},''
  \href{http://dx.doi.org/10.4310/ATMP.2014.v18.n2.a4}{{\em
  Adv.Theor.Math.Phys.} {\bfseries 18} (2014) 401--467},
\href{http://arxiv.org/abs/1306.0002}{{\ttfamily arXiv:1306.0002 [hep-th]}}.

\bibitem{Alim:2012ss}
M.~Alim and E.~Scheidegger, ``{Topological strings on elliptic fibrations},''
  \href{http://dx.doi.org/10.4310/CNTP.2014.v8.n4.a4}{{\em
  Commun.Num.Theor.Phys.} {\bfseries 08} (2014) 729--800},
\href{http://arxiv.org/abs/1205.1784}{{\ttfamily arXiv:1205.1784 [hep-th]}}.

\bibitem{Aganagic:2006ho}
M.~Aganagic, V.~Bouchard, and A.~Klemm, ``{Topological strings and (almost)
  modular forms},'' \href{http://dx.doi.org/10.1007/s00220-007-0383-3}{{\em
  Communications in Mathematical Physics} {\bfseries 277} no.~3, (Feb., 2008)
  771--819}.

\bibitem{Haghighat:2008gw}
B.~Haghighat, A.~Klemm, and M.~Rauch, ``{Integrability of the holomorphic
  anomaly equations},''
  \href{http://dx.doi.org/10.1088/1126-6708/2008/10/097}{{\em JHEP} {\bfseries
  0810} (2008) 097},
\href{http://arxiv.org/abs/0809.1674}{{\ttfamily arXiv:0809.1674 [hep-th]}}.

\bibitem{Marino:2007te}
M.~Mari\~no, R.~Schiappa, and M.~Weiss, ``{Nonperturbative effects and the
  large-order behavior of matrix models and topological strings},''
  \href{http://dx.doi.org/10.4310/CNTP.2008.v2.n2.a3}{{\em
  Commun.Num.Theor.Phys.} {\bfseries 2} (2008) 349--419},
\href{http://arxiv.org/abs/0711.1954}{{\ttfamily arXiv:0711.1954 [hep-th]}}.

\bibitem{Simon:1970aa}
B.~Simon, ``Coupling constant analyticity for the anharmonic oscillator (with
  an appendix by A. Dicke),'' {\em Ann. Phys.} {\bfseries 58} (1970) 76--136.

\bibitem{Bender:1973rz}
C.~M. Bender and T.~T. Wu, ``{Anharmonic oscillator II a study of pertubation
  theory in large order},''
  \href{http://dx.doi.org/10.1103/PhysRevD.7.1620}{{\em Phys.Rev.} {\bfseries
  D7} (1973) 1620--1636}.

\bibitem{e84}
J.~\'Ecalle, ``Cinq applications des fonctions r\'esurgentes,'' {\em Publ.
  Math. d'Orsay} {\bfseries 84-62} (1984) .

\bibitem{Polchinski:1994fq}
J.~Polchinski, ``{Combinatorics of boundaries in string theory},''
  \href{http://dx.doi.org/10.1103/PhysRevD.50.R6041}{{\em Phys. Rev.}
  {\bfseries D50} (1994) 6041--6045},
\href{http://arxiv.org/abs/hep-th/9407031}{{\ttfamily arXiv:hep-th/9407031
  [hep-th]}}.

\bibitem{Martinec:2003ka}
E.~J. Martinec, ``{The Annular report on noncritical string theory},''
\href{http://arxiv.org/abs/hep-th/0305148}{{\ttfamily arXiv:hep-th/0305148
  [hep-th]}}.

\bibitem{Alexandrov:2003nn}
S.~{\relax Yu}. Alexandrov, V.~A. Kazakov, and D.~Kutasov, ``{Nonperturbative
  effects in matrix models and D-branes},''
  \href{http://dx.doi.org/10.1088/1126-6708/2003/09/057}{{\em JHEP} {\bfseries
  09} (2003) 057},
\href{http://arxiv.org/abs/hep-th/0306177}{{\ttfamily arXiv:hep-th/0306177
  [hep-th]}}.

\bibitem{Marino:2006hs}
M.~Mari\~no, ``{Open string amplitudes and large order behavior in topological
  string theory},'' \href{http://dx.doi.org/10.1088/1126-6708/2008/03/060}{{\em
  JHEP} {\bfseries 03} (2008) 060},
\href{http://arxiv.org/abs/hep-th/0612127}{{\ttfamily arXiv:hep-th/0612127
  [hep-th]}}.

\bibitem{Pasquetti:2009jg}
S.~Pasquetti and R.~Schiappa, ``{Borel and Stokes nonperturbative phenomena in
  topological string theory and c=1 matrix models},''
  \href{http://dx.doi.org/10.1007/s00023-010-0044-5}{{\em Annales Henri
  Poincare} {\bfseries 11} (2010) 351--431},
\href{http://arxiv.org/abs/0907.4082}{{\ttfamily arXiv:0907.4082 [hep-th]}}.

\bibitem{Strominger:1990et}
A.~Strominger, ``{Heterotic solitons},''
\href{http://dx.doi.org/10.1016/0550-3213(90)90599-9}{{\em Nucl.Phys.}
  {\bfseries B343} (1990) 167--184}.

\bibitem{Callan:1991dj}
J.~Callan, Curtis~G., J.~A. Harvey, and A.~Strominger, ``{World sheet approach
  to heterotic instantons and solitons},''
\href{http://dx.doi.org/10.1016/0550-3213(91)90074-8}{{\em Nucl.Phys.}
  {\bfseries B359} (1991) 611--634}.

\bibitem{Callan:1991ky}
J.~Callan, Curtis~G., J.~A. Harvey, and A.~Strominger, ``{Worldbrane actions
  for string solitons},''
\href{http://dx.doi.org/10.1016/0550-3213(91)90041-U}{{\em Nucl.Phys.}
  {\bfseries B367} (1991) 60--82}.

\bibitem{Couso-Santamaria:2015aa}
R.~Couso-Santamar\'ia, M.~Mari\~no, and R.~Schiappa, work in progress.

\bibitem{Becker:1995kb}
K.~Becker, M.~Becker, and A.~Strominger, ``{Five-branes, membranes and
  nonperturbative string theory},''
  \href{http://dx.doi.org/10.1016/0550-3213(95)00487-1}{{\em Nucl.Phys.}
  {\bfseries B456} (1995) 130--152},
\href{http://arxiv.org/abs/hep-th/9507158}{{\ttfamily arXiv:hep-th/9507158
  [hep-th]}}.

\bibitem{Candelas:1990rm}
P.~Candelas, X.~C. De~La~Ossa, P.~S. Green, and L.~Parkes, ``{A pair of
  Calabi--Yau manifolds as an exactly soluble superconformal theory},''
\href{http://dx.doi.org/10.1016/0550-3213(91)90292-6}{{\em Nucl. Phys.}
  {\bfseries B359} (1991) 21--74}.

\bibitem{Strominger:1990pd}
A.~Strominger, ``{Special geometry},''
\href{http://dx.doi.org/10.1007/BF02096559}{{\em Commun. Math. Phys.}
  {\bfseries 133} (1990) 163--180}.

\bibitem{Pioline:2009ia}
B.~Pioline and S.~Vandoren, ``{Large D-instanton effects in string theory},''
  \href{http://dx.doi.org/10.1088/1126-6708/2009/07/008}{{\em JHEP} {\bfseries
  07} (2009) 008},
\href{http://arxiv.org/abs/0904.2303}{{\ttfamily arXiv:0904.2303 [hep-th]}}.

\end{thebibliography}\endgroup
